# Optimal Power Scheduling for High Renewables-Integrated Energy Systems with Battery Storage

Cunzhi Zhao *Member, IEEE* and Xingpeng Li, *Senior Member, IEEE*

***Abstract*— In high renewables-integrated power systems, irrespective to their sizes, energy storage is commonly included and utilized to mitigate fluctuations from both the load and renewable power generation, ensuring system reliability, among which battery energy storage system (BESS) are experiencing fast-growth in recent years. The BESS systems, predominantly employing lithium-ion batteries, have been extensively deployed. The degradation of these batteries significantly affects system efficiency. Deep neural networks can accurately quantify the battery degradation; however, the model complexity hinders their applications in energy scheduling for various power systems at different scales. To address this issue, this paper presents a novel approach, introducing a linearized sparse neural network-based battery degradation model (SNNBD), specifically tailored to quantify battery degradation based on the scheduled BESS operational profiles. This approach achieves accurate degradation prediction while substantially reducing the complexity associated with a dense neural network model. The computational burden of day-ahead energy scheduling when integrating battery degradation can thus be substantially alleviated. Case studies, conducted on both small-scale microgrids and large-scale bulk power grids, demonstrated the efficiency and suitability of the proposed optimal energy scheduling model that can effectively address battery degradation concerns while optimizing day-ahead energy scheduling operations.**

***Index Terms*— Battery degradation modeling, Bulk power grids, Day-ahead scheduling, Energy management, Machine learning, Microgrids, Optimization, Sparse neural network.**

## NOMENCLATURE

*Indices:*

| | |
|---|---|
| $g$ | Generator index. |
| $s$ | Battery energy storage system index. |
| $k$ | Transmission line index. |
| $l$ | Load index. |
| $wt$ | Wind turbine index. |
| $pv$ | Photovoltaic index. |

*Sets:*

| | |
|---|---|
| $T$ | Set of time intervals. |
| $G$ | Set of controllable micro generators. |
| $S$ | Set of energy storage systems. |
| $WT$ | Set of wind turbines. |
| $PV$ | Set of PV systems. |

*Parameters:*

| | |
|---|---|
| $c_g$ | Linear cost for controllable unit *g*. |
| $c_g^{NL}$ | No load cost for controllable unit *g*. |
| $c_g^{SU}$ | Start-up cost for controllable unit *g*. |
| $\Delta T$ | Length of a single dispatch interval. |
| $R_{prcnt}$ | Ratio of the backup power to the total power. |
| $E_s^{Max}$ | Maximum energy capacity of ESS *s*. |
| $E_s^{min}$ | Minimum energy capacity of ESS *s*. |
| $c_t^{Buy}$ | Wholesale electricity purchase price in time interval *t*. |
| $c_t^{Sell}$ | Wholesale electricity sell price in time interval *t*. |
| $P_g^{Max}$ | Maximum capacity of generator *g*. |
| $P_g^{Min}$ | Minimum capacity of generator *g*. |
| $P_k^{Max}$ | Maximum thermal limit of transmission line *k*. |
| $b_k$ | Susceptance, inverse of impedance, of branch *k*. |
| $P_{Grid}^{Max}$ | Maximum thermal limit of tie-line between main grid and microgrid. |
| $P_g^{Ramp}$ | Ramping limit of diesel generator *g*. |
| $P_s^{Max}$ | Maximum charge/discharge power of BESS *s*. |
| $P_s^{Min}$ | Minimum charge/discharge power of BESS *s*. |
| $\eta_s^{Disc}$ | Discharge efficiency of BESS *s*. |
| $\eta_s^{Char}$ | Charge efficiency of BESS *s*. |

*Variables:*

| | |
|---|---|
| $U_t^{Buy}$ | Status of buying power from main grid in time interval *t*. |
| $U_t^{Sell}$ | Status of selling power to main grid status in time *t*. |
| $U_{s,t}^{Char}$ | Charging status of energy storage system *s* in time interval *t*. It is 1 if charging status; otherwise 0. |
| $U_{s,t}^{Disc}$ | Discharging status of energy storage system *i* in time interval *t*. It is 1 if discharging status; otherwise 0. |
| $U_{g,t}$ | Status of generator *g* in time interval *t*. It is 1 if on status; otherwise 0. |
| $V_{g,t}$ | Startup indicator of Status of generator *g* in time interval *t*. It is 1 if unit g starts up; otherwise 0. |
| $P_{g,t}$ | Output of generator *g* in time interval *t*. |
| $\theta_{n(k)}^t$ | Phase angle of sending bus *n* of branch *k*. |
| $\theta_{m(k)}^t$ | Phase angle of receiving bus *m* of branch *k*. |
| $P_{k,t}$ | Line flow at transmission line *k* at time period *t*. |
| $P_t^{Buy}$ | Amount of power purchased from main grid power in time interval *t*. |
| $P_t^{Sell}$ | Amount of power sold to main grid power in time interval *t*. |
| $P_{l,t}$ | Demand of the microgrid in time interval *t*. |
| $P_{s,t}^{Disc}$ | Discharging power of energy storage system *s* at time *t*. |
| $P_{s,t}^{Char}$ | Charging power of energy storage system s at time t. |

## I. INTRODUCTION

Renewable energy sources (RES) have emerged as a pivotal component of the future power system, due to their environmental friendly attributes in contrast to conventional fossil fuels. By producing clean, sustainable, and inexhaustible electric energy, RES plays a transformative role in reducing greenhouse gas emissions in the electricity sector and thus mitigating climate change [1]. Nonetheless, the escalating utilization of RES for power generation has introduced inherent stability challenges in the system, primarily due to the unpredictable and intermittent nature of deeply integrated RES

Cunzhi Zhao is with Department of Engineering and Computer Science (ENCS) at McNeese State University; he was with the University of Houston. Xingpeng Li is with the Department of Electrical and Computer Engineering, University of Houston. (e-mail: czhao20@uh.edu; Xingpeng.Li@asu.edu).

[2]-[4]. In response to this challenge, battery energy storage systems (BESS) are being extensively adopted as an effective and practically viable solution [5]. BESS effectively addresses the variability and uncertainty inherent in RES by efficiently storing excess renewable energy during peak periods and releasing it during off-peak periods of renewable generation [6]. This capability not only promotes a seamless integration of renewable energy in the grid but also reinforces the resilience of the system. Furthermore, BESS plays a pivotal role in providing essential ancillary services such as frequency regulation, voltage control, and peak shaving, thereby enhancing the stability and efficiency of the overall power system [7]-[8].Numerous studies have demonstrated the successful integration of BESS into both bulk power systems and microgrids, particularly those integrating high penetrations of RES. For instance, [9]-[10] demonstrate the microgrid's ability to support the main grid with integrated BESS. Moreover, [11] highlights the significant benefits of incorporating BESS into the power system. Another notable example is the offshore BESS system presented in [12], which effectively reduces carbon emissions. Various other models have been proposed to incorporate BESS to mitigate fluctuations caused by renewable energy sources, as presented in [13]-[16]. In summary, the deployment of BESS is indispensable for the successful integration of renewable energy into the power system. It not only improves the system's stability and efficiency but also paves the way for a cleaner and more sustainable energy future.

Lithium-ion batteries are the dominant component in commercial BESS [17]. However, their chemical nature makes them prone to degradation over cycling, leading to performance deterioration and efficiency loss. Degradation mainly arises from Li-ion depletion, electrolyte breakdown, and internal resistance growth, which collectively reduce available energy capacity [19]-[20]. Aging is further influenced by ambient temperature, charge/discharge rate, state of charge (SOC), state of health (SOH), and depth of discharge (DOD) [21]-[22]. Accurately quantifying battery degradation under these varying factors remains a challenging but crucial task, especially as BESS operate under diverse conditions in both microgrids and bulk power systems. DOD-based battery degradation or linear degradation models exhibit limitations primarily due to their oversimplified approach, relying solely on the depth to which a battery is discharged. These models often lack the necessary detail to comprehensively capture the intricate interplay of various factors influencing battery health, such as SOC, ambient temperature, charge/discharge rate, and SOH. The resultant lack of precision in predicting battery degradation, especially over the long term, is a consequence of neglecting these crucial parameters. Additionally, the static nature of DOD-based models poses challenges in adapting to dynamic operating conditions and may not accurately reflect the complexities of real-world scenarios, particularly in systems with varied and unpredictable usage patterns.

Various battery degradation models have been developed, yet most fail to comprehensively capture degradation across diverse operating conditions. DOD-based models [23]-[27] and linear degradation models [28] offer only rough estimates, as they neglect key parameters such as SOC, temperature, and SOH, leading to limited prediction accuracy in day-ahead scheduling. To overcome these limitations, neural network–based battery degradation (NNBD) models have been proposed [29]. Unlike traditional models, NNBD considers multiple operational factors including SOC, DOD, temperature, charge/discharge rate, and SOH for each cycle, achieving more accurate degradation estimation. However, when multiple BESS units are involved, NNBD integration introduces significant computational complexity, as the number of constraints and variables increases exponentially, making the optimization problem difficult to solve within practical timeframes. This motivates the development of efficient sparse neural network structures that can retain accuracy while mitigating computational burden.

Several heuristic and compression techniques have been proposed to alleviate neural network complexity. For instance, pruning unimportant edges in neural belief propagation decoders [30] and sparse feature learning [31] help reduce computational demands, although sometimes at the expense of training accuracy. Sparse convolutional neural networks [32] and pruning-based simplifications have been applied in power amplifier models [33] and SCADA systems [34]. Furthermore, recent studies emphasize the issue of overparameterization in deep neural networks [35]-[36], suggesting that sparse model offers a practical approach to balancing computational feasibility and model precision.

Since the sparsity and pruning techniques have proved to be efficient to reduce the complexity of neural networks in many other applications, it may be a perfect solution to obtain a low computational complexity model in battery degradation prediction. Thus, we propose a sparse neural network-based battery degradation model (SNNBD) to quantify the battery degradation in BESS daily operations. SNNBD is designed to be significantly less complex than the traditional fully-connected dense neural network model. SNNBD is designed to reduce the computation burden induced by the rectified linear unit (ReLU) activation function. Achieving this entails a strategic process of pruning during training, whereby a predetermined percentage of neurons is systematically pruned. The sparsity percentage is defined as the ratio of pruned neurons to the total neurons in the neural network. A higher percentage of sparsity may decrease the computation complexity significantly, but the accuracy of the battery degradation prediction may decrease as compared with a less-sparse or dense model. It will be a trade-off between the sparsity and the training accuracy. Compared to the original NNBD model, the proposed SNNBD model contains only a percentage of NNBD's neurons which may reduce the computational burden significantly while maintaining accurate battery degradation prediction.

The main contributions of this paper are as follows:

- *Refined Battery Degradation Modeling:* The proposed SNNBD model significantly refines NNBD model, elevating its proficiency in quantifying battery degradation within the day-ahead scheduling model.
- *Computational Augmentation with SNNBD:* To efficiently address the day-ahead scheduling optimization challenge, this paper proposes an innovative SNNBD-assisted computational enhancement model. Capitalizing on the capabilities of the SNNBD model, this enhancement substantially improves the computational efficiency of the optimization process. This, in turn, translates into more responsive and informed decision-making procedures.

- *Linearization Technique for Practicality:* The integration of the SNNBD model into the day-ahead scheduling framework is accompanied by a pertinent linearization technique. This technique simplifies the model's analysis and evaluation, making it more practical and feasible for real-world application scenarios.
- *In-depth Economic Insights:* This paper provides an insightful market analysis. By comparing locational marginal prices (LMPs) across three scenarios: (1) zero-degradation BESS, (2) degraded BESS, and (3) no BESS integration, the economic implications and advantages of incorporating BESS into the power system and capturing its degradation are explored. This analysis provides a comprehension of the economic landscape, enriching decision-making processes within the energy market.

The rest of the paper is organized as follows. Section II describes the sparse neural network model and training strategy. Section III presents the traditional day-ahead scheduling model. Section IV presents the SNNBD integrated day-ahead scheduling model. Section V presents case studies and Section VI concludes the paper.

## II. Sparse Neural Network Based Battery Degradation Model

This section outlines the training process for the proposed SNNBD model. We proposed two training schemes: (i) Warm Start that trains the SNNBD based on the pre-trained NNBD model, and (ii) Cold Start that trains the SNNBD model directly with random initial weights. Both models consist of 5 input neurons, 20 neurons in hidden layer 1, 10 neurons in hidden layer 2, and 1 neuron in the output layer. The hidden layers utilize the ReLU as the activation function for each neuron.

### A. NNBD Model

The NNBD model was trained using battery aging data generated in MATLAB/Simulink [37], where a battery model with a cycle generator simulates charging and discharging processes under varying SOC, DOD, ambient temperatures, and current rates to replicate diverse operating conditions. A grid search was applied to identify optimal hyperparameters. Training used a batch size of 256 and an adaptive learning rate. The batch size test results are summarized in Table I.

Table I Batch size tests

| Batch-Size | 32 | 64 | 128 | 256 | 512 |
|---|---|---|---|---|---|
| Accuracy | 50% | 90.5% | 94.0% | 94.5% | 92.5% |

### B. Warm Start

The training process for Warm Start is illustrated in the algorithm explained below. The training algorithm for the SNNBD leveraging a pre-trained NNBD model starts by initializing parameters such as the sparsity level $\varepsilon$ and the number of training epochs. It then iterates through each epoch, beginning with calculating a threshold value based on $\varepsilon$ to identify which weights should be pruned. This thresholding step is crucial as it determines the extent of sparsity introduced into the network. Subsequently, a binary mask matrix $\sigma$ is created based on this threshold, effectively zeroing out weights below the threshold. The weights matrices $\omega$ are then updated by element-wise multiplication with the mask $\sigma$, preserving only the weights deemed significant for the network's performance. The weights in each layer are sorted by their absolute values, and the weights with the smallest magnitudes are set to zero by applying a masking operation. Pruning the smallest magnitude neurons helps to recover from any pruning-induced loss in accuracy. Following this pruning step, the remaining weights are fine-tuned using gradient descent or another optimization algorithm to adjust to the new sparsity pattern and maintain or improve the network's predictive accuracy. This iterative process of thresholding, pruning, and fine-tuning allows the network to adapt to different levels of sparsity, offering a balance between computational efficiency and model performance.

**Training Algorithm: Sparse Neural Network**

1. Obtain the weights matrices $\omega$ from the pre-trained NNBD model.
2. Set the sparse percentage $\varepsilon$.
3. Set the training epochs.
4. **For** $e$ *in* epochs,
5. | Find the threshold weights with sparse percentage $\varepsilon$.
6. | Set the Pruning masks matrices $\sigma$ with the threshold weights.
7. | Update weights matrices $\boldsymbol{\omega}'$ by $\boldsymbol{\omega} * \sigma$.
8. | Tune the weights matrices with gradient descent.
9. **end For**

### C. Cold Start

Cold Start offers a simple approach compared to Warm Start. Instead of training the neural network based on the fine-tuned NNBD weights, Cold Start directly trains a sparse neural network using random initial weights. In essence, the key difference between Warm Start and Cold Start lies in the choice of initial weights. However, all other training techniques remain consistent between the two options. The performance and efficiency of both Warm Start and Cold Start will be evaluated and compared.

### D. Pruning Method

Pruning is a technique employed in neural networks to reduce the size and complexity of a model by eliminating unnecessary connections or neurons [38]. The objective of pruning is to enhance the efficiency of the training model, minimize memory requirements, and potentially improve its generalization capabilities. In neural network optimization, pruning within sparse neural networks and dropout employed in conventional neural network training serve distinct purposes. Pruning aims to enhance network efficiency by systematically eliminating less critical neurons and connections, resulting in a more compact and resource-efficient architecture as shown in Fig. 1.

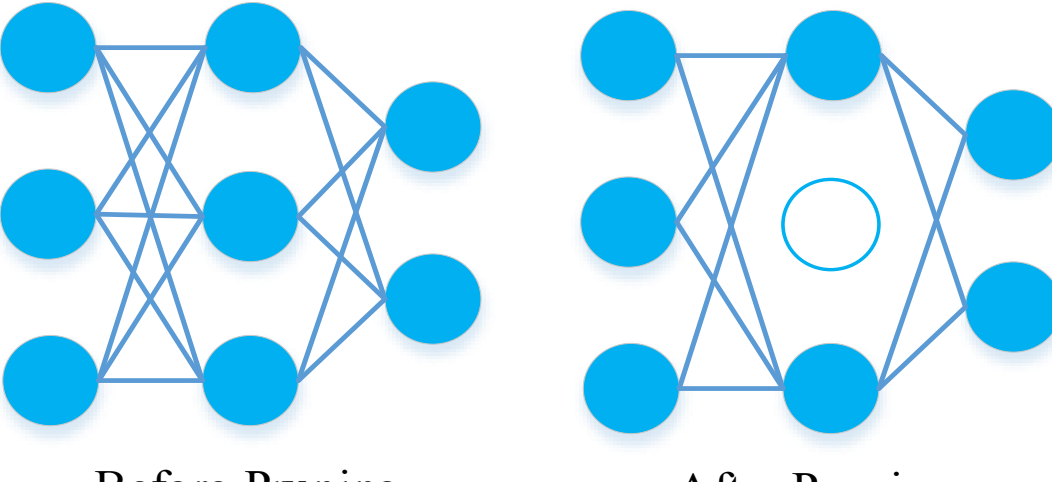

Fig. 1. Pruning of a sample neural network model.

The pruning masks are regenerated for each epoch, ensuring that they remain consistent throughout the training process. For weights that is not a bias from each training epoch, it creates a matrix copy of its absolute values and calculates a pruning threshold based on the specified percentile of weight magnitudes. Using this threshold, a binary mask (pruning mask) is generated, where values above the threshold are set to 1 and others to 0. The pruning mask share the same dimension with the weight matrix of the neural network. Finally, the original parameter is element-wise multiplied by this pruning mask, effectively pruning weights below the threshold to zero as shown in Fig. 2.

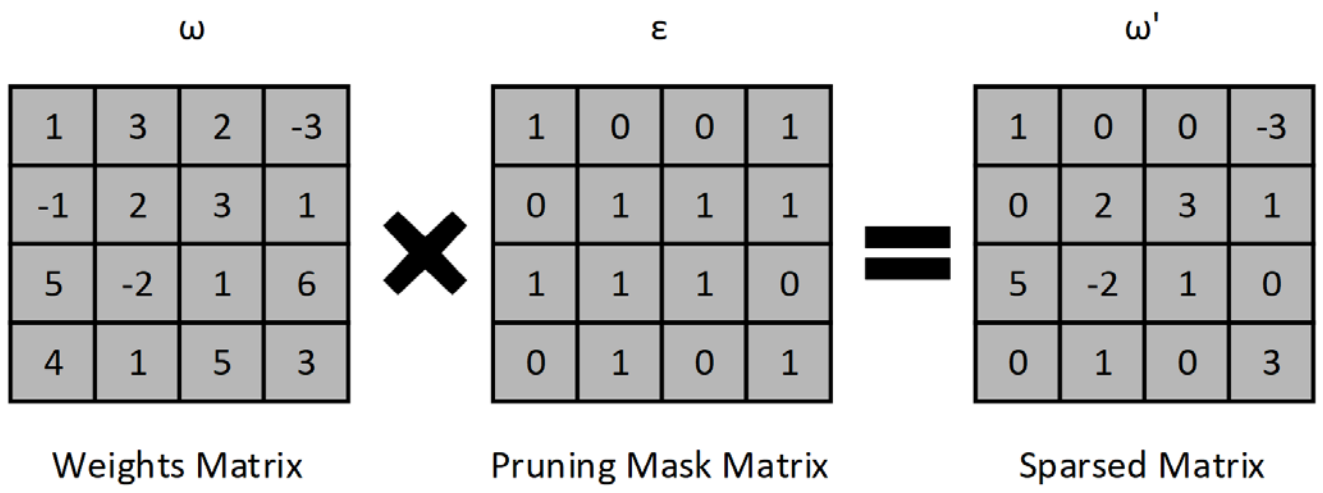

Fig. 2. Example of pruning process.

This process is performed in a no-gradient context, making it suitable for post-training weight pruning without affecting the model's training process. This consistency enhances the robustness of the proposed SNNBD model. These pruning masks play a crucial role in achieving a compact representation of the sparse neural network. Instead of storing and computing all connection weights, only the active connections are considered. This approach significantly reduces both memory usage and computational demands. The integration of pruning masks allows sparse neural networks to strike a balance between model complexity and efficiency. This feature makes them particularly valuable for a variety of applications, especially in situations with limited computational resources or deployment constraints. Sparse neural networks, by selectively retaining important connections and eliminating less relevant ones, offer an effective and resource-efficient solution for real-world scenarios.

### *E. Fine Tuning and Setup*

After the pruning stage, the network undergoes retraining to restore and fine-tune its performance in the next epoch. During retraining, the pruning mask plays a crucial role in removing the pruned connections, effectively fixing their weights at zero. Only the remaining active connections are updated during the retraining process. This allows the network to relearn and redistribute its capacity among the remaining connections, compensating for the pruned ones. The sparse neural network is trained using mini-batch gradient descent, with performance evaluated by the mean squared error (MSE) loss function shown in equation (1), which measures the average squared difference between predicted and actual values.

$$Mean\ Square\ Error = \frac{1}{n}\sum_{i=1}^{n}(y_i - \tilde{y_i})^2 \tag{1}$$

## III. Traditional Day-Ahead Scheduling Model

This section presents the day-ahead scheduling problem for both the bulk power system and the microgrid system. It's important to note that neither of the models listed in this section consider the battery degradation.

### *A. Bulk Power System Energy Scheduling Model*

The day-ahead scheduling problem of the bulk power system is represented by the traditional security constrained unit commitment (SCUC) model. The objective of the traditional SCUC model is to minimize the total operating cost of the system as defined in equation (2).

Objective function:

$$f^{Cost} = \sum\sum P_{g,t}c_{g,t} + U_g c_g^{NL} + V_g c_g^{SU}, \forall g,t \tag{2}$$

Constraints:

$$\sum_{g\in S_G} P_{g,t} + \sum_{wt\in S_{WT}} P_{wt,t} + \sum_{pv\in S_{PV}} P_{pv,t} + \sum_{s\in S_S} P_{s,t}^{Disc} + \sum_{k\in S_{n-}} P_{k,t} = \sum_{k\in S_{n+}} P_{k,t} \sum_{l\in S_L} P_{l,t} + \sum_{s\in S_S} P_{s,t}^{Char} \tag{3}$$

$$P_g^{Min} \le P_{g,t} \le P_g^{Max} \ ,\forall g,t \tag{4}$$

$$P_{g,t+1} - P_{g,t} \le \Delta T \cdot P_g^{Ramp}, \forall g,t \tag{5}$$

$$P_{g,t} - P_{g,t+1} \le \Delta T \cdot P_g^{Ramp}, \forall g,t \tag{6}$$

$$V_{g,t} \ge U_{g,t} - U_{g,t-1}, \forall g,t, \tag{7}$$

$$V_{g,t+1} \le 1 - U_{g,t}, \forall g,t, \tag{8}$$

$$V_{g,t} \le U_{g,t}, \forall g,t, \tag{9}$$

$$-P_k^{Max} \le P_{k,t} \le P_k^{Max} \ ,\forall k,t, \tag{10}$$

$$P_{k,t} - b_k\left(\theta_{n(k)}^t - \theta_{m(k)}^t\right) = 0 \ ,\forall k,t, \tag{11}$$

$$U_{s,t}^{Disc} + U_{s,t}^{Char} \le 1, \forall s,t \tag{12}$$

$$U_{s,t}^{Char} \cdot P_s^{Min} \le P_{s,t}^{Char} \le U_{s,t}^{Char} \cdot P_s^{Max}, \forall s,t \tag{13}$$

$$U_{s,t}^{Disc} \cdot P_s^{Min} \le P_{s,t}^{Disc} \le U_{s,t}^{Disc} \cdot P_s^{Max}, \forall s,t \tag{14}$$

$$E_{s,t} - E_{s,t-1} + \Delta T\left(P_{s,t-1}^{Disc}/\eta_s^{Disc} - P_{s,t-1}^{Char}\eta_s^{Char}\right) = 0, \forall s,t \tag{15}$$

$$E_{s,t=24} = E_s^{Initial}, \forall s \tag{16}$$

$$0 \le E_{s,t} \le E_{s,t}^{max} \tag{17}$$

The power balance equation for bus n incorporates synchronous generators, renewable energy sources, battery energy storage systems, and load demand, as represented by equation (3). Constraints (4-6) define the power output limits and ramping limits for each generator. To establish the relationship between a generator's start-up status and its on/off status, equations (7)-(9) are employed. Equation (10) enforces the thermal limit of the transmission lines. Constraint (11) calculates the power flow within the network.

For the BESS, the state of charge (SOC) level is determined by the ratio between the current stored energy and the maximum available energy capacity, as shown in equation (12). Constraints (13)-(14) maintain the charging/discharging power of the BESS within specified limits. Equation (15) calculates the stored energy of the BESS for each time interval. Equation (16) mandates that the final SOC level of the BESS matches the initial value. Equation (17) establishes the upper limit for the stored energy of the BESS.

### *B. Microgrid Energy Scheduling Model*

The traditional microgrid day-ahead scheduling problem shares some constraints of the bulk power system model, excluding the power flow constraints. The objective function for microgrids aims to minimize the total cost, incorporating the cost of traditional generators and the cost of tie-line power exchange, as depicted in (18).

Objective function:

$$f^{Cost} = \sum\sum(P_{g,t}c_{g,t} + U_g c_g^{NL} + V_g c_g^{SU}) + P_t^{Buy}c_t^{Buy} - P_t^{Sell}c_t^{Sell}, \forall g,t \quad (18)$$

<u>*Constraints*</u>:

The power balance equation for microgrid is presented in (19). To ensure the appropriate status of power exchange between the microgrid and the main grid, (20) is utilized, specifying the status of being a buyer, seller, or idle. Constraints (21) and (22) limit the thermal limits of the tie-line. Lastly, equation (23) setup the emergency reserve of the system. The traditional microgrid day-ahead scheduling constraints encompass (4)-(9) and (12)-(23). Unlike the power flow constraints present in the bulk power system model, the microgrid model incorporates tie-line exchange equations within the day-ahead scheduling framework.

$$P_t^{Buy} + \sum_{g\in S_G} P_{g,t} + \sum_{wt\in S_{WT}} P_{wt,t} + \sum_{pv\in S_{PV}} P_{pv,t} + \sum_{s\in S_S} P_{s,t}^{Disc} = P_t^{Sell} + \sum_{l\in S_L} P_{l,t} + \sum_{s\in S_S} P_{s,t}^{Char} \quad (19)$$

$$U_t^{Buy} + U_t^{Sell} \leq 1, \forall t \quad (20)$$

$$0 \leq P_t^{Buy} \leq U_t^{Buy} \cdot P_{Grid}^{Max}, \forall t \quad (21)$$

$$0 \leq P_t^{Sell} \leq U_t^{Sell} \cdot P_{Grid}^{Max}, \forall t \quad (22)$$

$$P_{Grid}^{Max} - P_t^{Buy} + P_t^{Sell} + \sum_{g\in S_G}\left(P_g^{Max} - P_{g,t}\right) \geq R_{percent}\left(\sum_{l\in S_L} P_{l,t}\right), \forall t \quad (23)$$

## IV. Battery Degradation-Considered Day-Ahead Scheduling Model

### *A. Neural Network-Based Battery Degradation Quantification*

The battery degradation quantification is achieved by the SNNBD model. In this model, various variables of the BESS are processed and fed into the SNNBD model to predict the battery degradation per cycle of BESS operation. The input variables for the SNNBD model consist of temperature, SOC, DOD, C rate, and SOH. To calculate the DOD and C rate, (24) and (25) are utilized. Equation (26) predicts the battery degradation value, while (27) provides the equivalent degradation cost. In these equations, $c_{BESS}^{Capital}$ represents the capital investment cost of the BESS, $c_{BESS}^{SV}$ denotes the salvage value of the BESS, and $SOH_{EOL}$ signifies the state of health value considered as the end of battery life.

Consequently, the equivalent battery degradation cost is included as an additional component in the objective functions of traditional models. To model the day-ahead scheduling considering battery degradation, constraints (24)-(27) are added to the traditional day-ahead scheduling model. However, solving this battery degradation model directly is challenging due to the highly non-linear nature of the ReLU activation function used in the neural network.

$$\Delta DOD_t = |SOC_t - SOC_{t-1}| \quad (24)$$

$$C_t^{Rate} = \Delta DOD_t/\Delta T \quad (25)$$

$$BD = \sum_t SNNBD(Variables) \quad (26)$$

$$f^{BESS} = \frac{c_{BESS}^{Capital} - c_{BESS}^{SV}}{1 - SOH_{EOL}} BD \quad (27)$$

### *B. Piecewise-Linearization of SNNBD Model*

The SNNBD model is characterized by a series of equations that describe the calculation and activation processes of neurons. Each neuron's computation is defined by (28), which takes into account the input features from the first layer, the corresponding weight matrix $W$, and the biases matrix. The activation function is employed in the SNNBD model, as shown in (29). This activation function is commonly used in neural networks to introduce non-linearity and capture complex relationships between variables. However, the non-linearity of the ReLU function can pose challenges when it is embedded in the optimization problem. To address this issue, the ReLU activation function can be linearized using an auxiliary binary variable, denoted as $\delta_h^i$. This variable is further defined in (30)-(33). Specifically, $\delta_h^i$ takes a value of one when activation is enabled, meaning the input value is positive, and zero otherwise. The activated value of $x_h^i$, denoted as $a_h^i$, represents the output of the neuron after the ReLU activation is applied. In the formulation, theoretical bounds are applied based on the observed range of trained weights and biases. All weights and biases fall within [−2, 2]; therefore, a conservative big-M value of 10 is assigned to all neurons.

$$x_h^i = \sum x_{h-1}^i * W + Bias \quad (28)$$

$$a_h^i = relu(x_h^i) = max\,(0, x_h^i) \quad (29)$$

$$a_h^i \leq x_h^i + BigM * (1 - \delta_h^i) \quad (30)$$

$$a_h^i \geq x_h^i \quad (31)$$

$$a_h^i \leq BigM * \delta_h^i \quad (32)$$

$$a_h^i \geq 0 \quad (33)$$

The SNNBD-integrated day-ahead energy scheduling models for different systems are shown in the Table II.

Table II The proposed day-ahead scheduling models

| Systems | Equations |
|---|---|
| Bulk Power Grid | (2)-(17), (24)-(33) |
| Microgrid | (4)-(9), (12)-(33) |

### *C. Benchmark Model*

To evaluate the performance of the SNNBD model in day-ahead scheduling problems, a benchmark model will be employed. The benchmark model utilizes the NNBD model, which has been previously introduced in section II. The purpose of this benchmark model is to provide a basis for comparison and assess the effectiveness of the SNNBD model.

Both models are applied within the same day-ahead scheduling framework, sharing the same variables and constraints. The distinction between the two models lies in the methodology used to quantify battery degradation. By comparing the performance of the SNNBD model against the benchmark NNBD model, it becomes possible to evaluate the effectiveness of the sparse architecture introduced in the SNNBD model. This comparison aids in determining the advancements made by the SNNBD model in compact the neutral network structure, which in turn can contribute to reduce the computational complexity of the neural network integrated day-ahead scheduling problems.

# V. Case Studies

## A. Training Strategies: Warm Start vs. Cold Start

The dataset used in this study contains 3,628,782 data samples processed from 945 simulated battery aging tests, which were randomly divided into 80% for training and 20% for validation to ensure balanced and unbiased evaluation. The SNNBD model was trained using the Adam optimization algorithm with identical hyperparameters for both cold start and warm start approaches. The learning rate was initially set to 0.0001, with a batch size of 256, and a learning rate scheduler was applied with a step size of 20 epochs and a decay factor of 0.8 to enhance convergence and prevent overfitting.

The performance of cold start and warm start models was evaluated at different sparsity levels (10%, 20%, and 30%). In the cold start approach, the sparse network is trained from scratch with randomly initialized weights, whereas in the warm start approach, the sparse network is initialized using the weights of the fully trained 0% sparsity model. Table III summarizes the results in terms of MAE, RMSE, $R^2$, accuracy within different error tolerances (⩽5%, ⩽10%, ⩽15%), and the number of training epochs. Warm start consistently achieves lower MAE and comparable or slightly lower RMSE, as well as slightly higher $R^2$ in some cases, compared to cold start across all sparsity levels. Notably, warm start also demonstrates higher predictive accuracy: for example, the proportion of predictions within 5% error improves by 8–12%, while the proportions within 10% and 15% error show modest gains of 2–6% relative to cold start. These results indicate that initializing from a well-trained dense model provides better starting weights, enabling the sparse network to converge more effectively and achieve more accurate predictions. The benefits of warm start are especially pronounced at higher sparsity levels, where cold start struggles due to limited capacity and random initialization. Overall, these findings support the use of warm start for efficiently training sparse neural networks while maintaining high predictive accuracy.

Table III Results between Warm Start and Cold Start

| Sparsity | 10% | | 20% | | 30% | |
|---|---|---|---|---|---|---|
| | Cold Start | Warm Start | Cold Start | Warm Start | Cold Start | Warm Start |
| MAE | 0.0269 | 0.0247 | 0.0302 | 0.0247 | 0.0289 | 0.0249 |
| RMSE | 0.0413 | 0.0426 | 0.0503 | 0.0426 | 0.0447 | 0.0422 |
| $R^2$ | 0.9823 | 0.9812 | 0.9737 | 0.9811 | 0.9792 | 0.9815 |
| ≤5% | 59% | 67% | 58% | 66% | 53% | 65% |
| ≤10% | 88% | 90% | 86% | 90% | 84% | 90% |
| ≤15% | 93% | 94% | 93% | 94% | 92% | 94% |
| Epochs | 300 | 375 | 300 | 375 | 300 | 375 |

However, it's noteworthy that Cold Start requires fewer epochs to complete the training process. It is important to mention that the training epochs for Warm Start represent the combined training epochs required by the NNBD model and the SNNBD model, while for Cold Start, it refers to the training epochs of the sparse neural network alone. Training the sparse neural network from random initial weights (Cold Start) proves to be notably challenging when it comes to achieve an equivalently level of accuracy as the Warm Start. In contrast, Warm Start is designed to take advantage of the pre- trained NNBD model, which serves as a stable starting point. The SNNBD model is then applied to further refine and sparse the structure of the already trained model. This suggests that the pre-trained NNBD model provides a beneficial foundation for the SNNBD model. The initial training with the NNBD model establishes a solid baseline, and the subsequent application of the SNNBD model enables fine-tuning with sparsity. By leveraging the existing knowledge encoded in the NNBD model, Warm Start demonstrates superior training accuracy compared to Cold Start.

## B. SNNBD Model Training

All the results presented here are based on training Warm Start, as it outperforms the Cold Start. The training results of the proposed SNNBD model are depicted in Fig. 3 and Table IV. In Fig. 3, the 0% sparsity represents the original NNBD model without any sparsity applied. The subsequent markers—5%, 10%, and 15%—tinted in blue, red, and green, respectively, signify distinct error tolerance thresholds. Notably, the pattern that unfolds the interplay between sparsity and prediction accuracy.

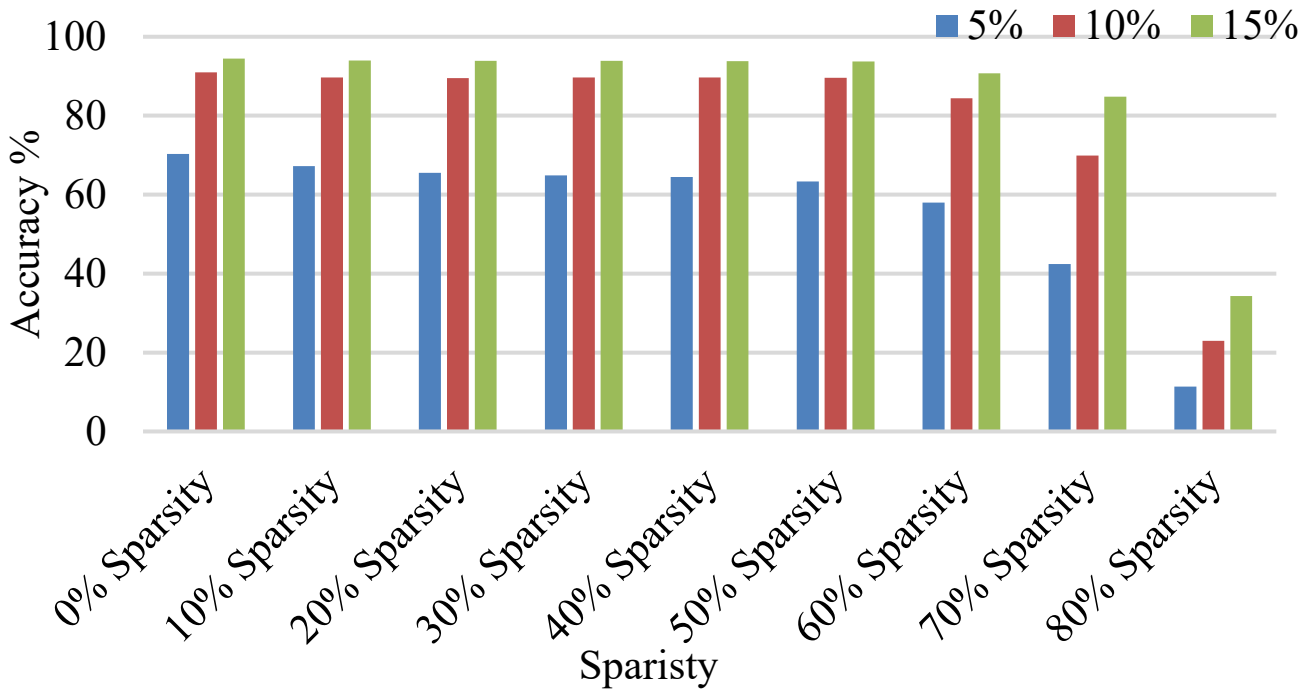

Fig. 3. Training results of SNNBD model at different sparsity levels.

Table IV. Training results of proposed SNNBD model under different sparsity levels and error tolerances

| Sparsity (%) | MAE | RMSE | $R^2$ | Error Tolerances | | |
|---|---|---|---|---|---|---|
| | | | | 5% | 10% | 15% |
| 0 | 0.024695 | 0.042643 | 0.981122 | 70% | 91% | 95% |
| 10 | 0.024708 | 0.042611 | 0.981151 | 67% | 90% | 94% |
| 20 | 0.024735 | 0.042619 | 0.981144 | 66% | 90% | 94% |
| 30 | 0.02489 | 0.042205 | 0.981509 | 65% | 90% | 94% |
| 40 | 0.024142 | 0.040621 | 0.98287 | 65% | 90% | 94% |
| 50 | 0.025461 | 0.041008 | 0.982542 | 63% | 90% | 94% |
| 60 | 0.027611 | 0.043288 | 0.980547 | 58% | 84% | 91% |
| 70 | 0.036276 | 0.052457 | 0.971434 | 42% | 70% | 85% |
| 80 | 0.19919 | 0.286237 | 0.149453 | 11% | 23% | 34% |

As sparsity percentage scales up, the precision of battery degradation value predictions undergoes a gradual decrement across all tolerance thresholds. This trend continues until the 70% sparsity mark is attained. When comparing the 0% sparsity model (NNBD model) and the 50% sparsity model, the accuracy stands at 94.5% and 93.7% respectively, considering a 15% error tolerance. The results also show that moderate sparsity improves model generalization by removing redundant connections without compromising accuracy. From 0% to 50% sparsity, both MAE and RMSE remain consistently low ( ≈ 0.024 – 0.026) and $R^2$ stays above 0.982, indicating that the sparse models maintain nearly the same predictive capability as the fully connected network. The best overall performance is observed around 40% sparsity, where the model achieves the lowest RMSE (0.0406) and highest $R^2$ (0.9829). Nevertheless, the model with 50% sparsity was selected for the subsequent case studies because it offers a more favorable trade-off between accuracy and computational efficiency, achieving comparable performance ($R^2$ = 0.9825). However, the 50% sparsity model significantly reduces the computational complexity compared to the original NNBD model since half of the neurons are pruned to be zero, thereby eliminating their connections. This reduction in computational complexity is exponential, as all connections associated with zero-valued neurons are discarded.

### *C. Microgrid Test Case*

To evaluate the performance of the integrated SNNBD day-ahead scheduling model, a typical grid-connected microgrid with renewable energy sources was employed as a testbed, as demonstrated in Section IV. The microgrid configuration consists of several components, including a traditional diesel generator, wind turbines, residential houses equipped with solar panels, and a lithium-ion BESS with a charging/discharging efficiency of 90%. The parameters for these main components are provided in Table V.

To simulate realistic conditions, the load data for the microgrid is based on the electricity consumption of 1000 residential houses. The ambient temperature and available solar power for a 24-hour period are sourced from the Pecan Street Dataport [39], ensuring accurate representation of real-world environmental conditions. The wholesale electricity price data is obtained from ERCOT [40], allowing the model to consider market dynamics in the day-ahead scheduling decisions.

The optimization problem, formulated as part of the day-ahead scheduling model, was solved on a computer with the following specifications: an AMD® Ryzen 7 3800X processor, 32 GB RAM, and an Nvidia Quadro RTX 2700 Super GPU with 8 GB of memory. The Pyomo [41] package, a powerful optimization modeling framework, was utilized to formulate and solve the day-ahead optimization problem. A high-performance mathematical programming solver Gurobi [42] was employed to efficiently find optimal solutions. By utilizing this realistic microgrid test platform and the computational resources mentioned, the SNNBD integrated day-ahead scheduling model can accurately capture the dynamics of the renewable energy sources, optimize the scheduling decisions, and assess the performance of the proposed approach.

Table V. Microgrid testbed

| Main Components | Diesel Generator | Wind Turbines | Solar Panels | Lithium-ion BESS |
|---|---|---|---|---|
| Size | 180kW | 1000kW | 1500kW | 300kWh |

Table VI presents the validation results for different sparsity levels of the SNNBD models in the microgrid day-ahead scheduling problem. The table provides insights into the performance of these models across various metrics. "Pseudo Total" represents the total cost with the SNNBD model, which serves as the objective of the day-ahead scheduling including the operating cost and degradation cost in optimization problem. "BD Cost" represents the equivalent battery degradation cost estimated using the SNNBD model. "Operation" shows the microgrid operating cost, including the cost associated with generators and power trading. "OG BD Cost" indicates the battery degradation cost obtained from the original NNBD model, which does not incorporate sparsity. "Updated Total" represents the sum of the operation cost and the "OG BD Cost". "0% sparsity" is considered as the benchmark model, used to evaluate the performance of the other SNNBD models with different sparsity levels.

From the information in Table VI, the solving time of the microgrid model decreases noticeably as the sparsity level increases from 0% to 40%. Although minor fluctuations are observed (e.g., the solving time at 20% sparsity is slightly higher than that at 10%), the overall trend clearly indicates that the SNNBD model improves computational efficiency compared to the dense model. In terms of system performance, the total cost and updated total cost remain nearly consistent across different SNNBD models and the benchmark model, indicating that the introduction of sparsity effectively reduces computation time without compromising optimization accuracy. These findings confirm that the sparse neural network structure achieves a favorable balance between computational efficiency and solution quality in the microgrid day-ahead scheduling problem.

Table VI. Microgrid energy scheduling results

| Sparsity | 0% | 10% | 20% | 30% | 40% |
|---|---|---|---|---|---|
| Pseudo Total ($) | 501.5 | 499.94 | 500.88 | 501.2 | 501.04 |
| BD Cost ($) | 7.27 | 7.93 | 8.85 | 7.4 | 8.22 |
| Operation ($) | 494.23 | 492.01 | 492.03 | 493.8 | 492.82 |
| OG BD Cost ($) | 7.27 | 11.25 | 11.2 | 8.09 | 9.27 |
| Updated Total ($) | 501.5 | 503.26 | 503.23 | 501.89 | 502.09 |
| Solving time (s) | 24.31 | 4.03 | 4.31 | 3.00 | 2.24 |

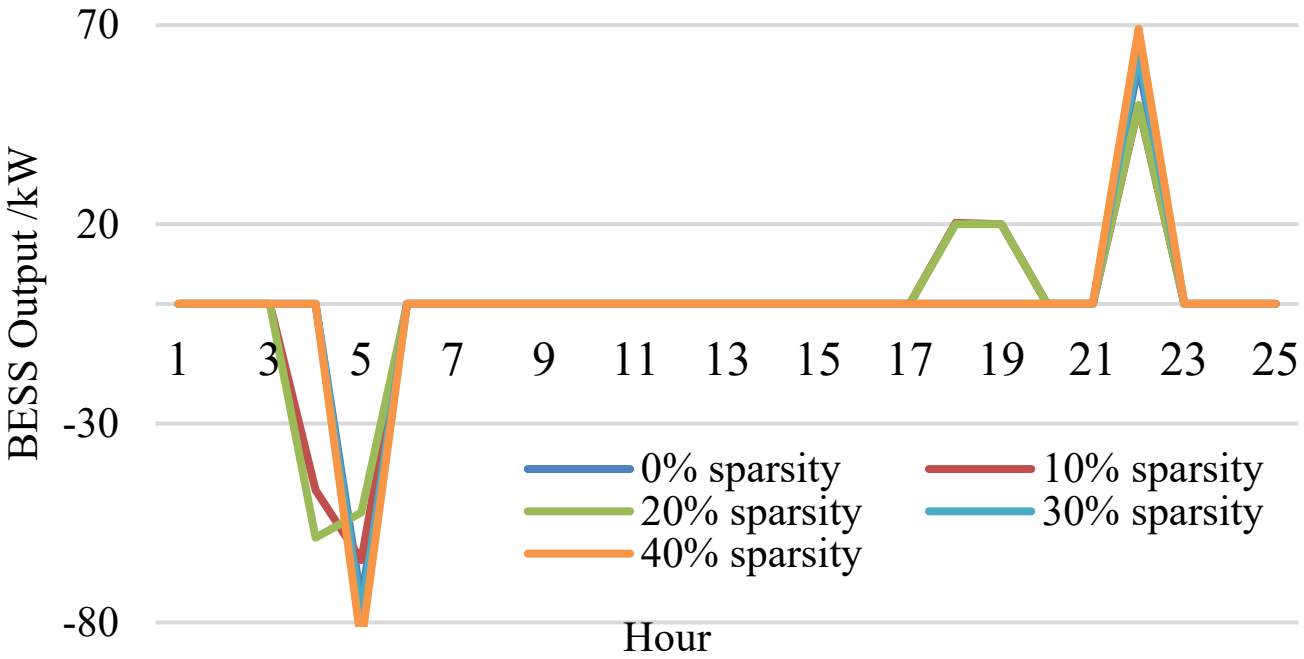

Fig. 4. BESS output in a microgrid system.

Fig. 4 illustrates the output curves of the BESS under different battery degradation models. The figure shows that the BESS charge/discharge power profiles largely overlap across most time intervals. The only notable difference is observed in the 10% and 20% sparsity models, where the BESS charges at 20 kW during the 7-8 pm. Overall, these results demonstrate that the SNNBD model is capable of finding solutions for the day-ahead scheduling problem. Based on these findings, it can be concluded that the SNNBD model is reliable and able to identify optimal solutions compared to the non-sparse NNBD model in the microgrid day-ahead scheduling problem. However, it should be noted that the SNNBD model does not yield efficiency improvements, even with higher sparsity levels. One possible reason for this observation could be the small scale of the microgrid case and the presence of only one BESS, which does not impose a heavy computational burden.

### *D. Bulk Power System Test Case*

To evaluate the day-ahead scheduling of the bulk power grid model, a typical IEEE 24-bus system (Fig. 5) is employed as a test bed. This system consists of 33 generators and serves as a representative model for large-scale power grids. In addition to the existing infrastructure, the test bed incorporates several BESSs and wind farms to evaluate their impact on the day-ahead scheduling. Fig. 5 illustrates the layout of the IEEE 24-bus system, showcasing the interconnected buses and the corresponding transmission lines. The objective of this evaluation is to optimize the scheduling decisions considering the presence of the multiple BESS and wind farm within the larger power grid system.

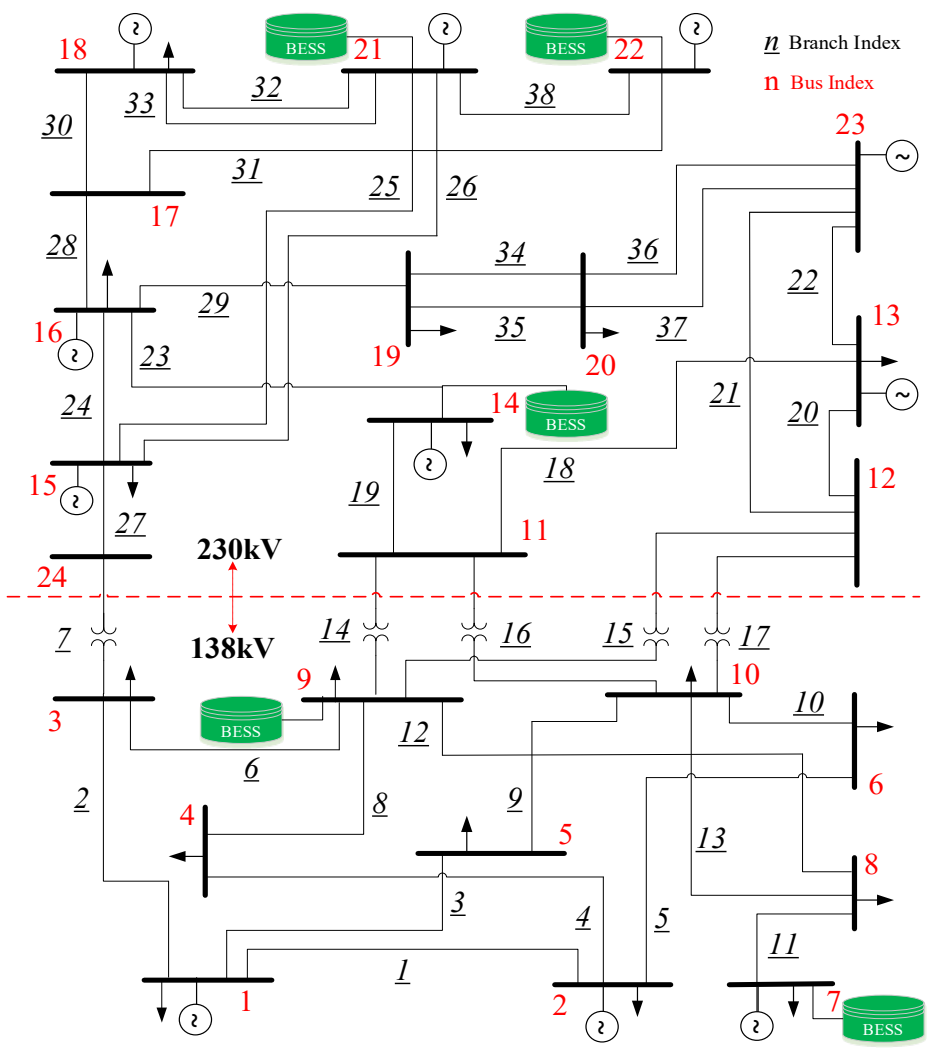

Fig. 5. Illustration of the modified IEEE 24-bus system.

Table VII provides the parameters of the BESSs installed at different buses within the IEEE 24-bus system. These parameters characterize the specifications of each BESS, including their energy capacities and power output capabilities. Notably, BESS number four possesses the largest energy capacity and the highest output power among the five BESSs considered in the system. The minimum power for charging or discharging is set to zero. Additionally, the IEEE 24-bus system incorporates five wind farms, each comprising a varying number of wind turbines. The capacity of each wind turbine is fixed at 200 kW. To obtain suitable wind profiles for this study, the wind profile data sourced from the Pecan Street Dataport [41] are appropriately scaled. The inclusion of these parameters and data in the evaluation allows for a comprehensive analysis of the day-ahead scheduling problem within the IEEE 24-bus system.

Table VII Setting of BESSs

| BESS No. | Bus No. | Energy Capacity (MWh) | Power Rating (MW) | Initial SOC |
|---|---|---|---|---|
| 1 | 21 | 50 | 20 | 40% |
| 2 | 22 | 10 | 4 | 40% |
| 3 | 7 | 10 | 4 | 40% |
| 4 | 14 | 200 | 100 | 40% |
| 5 | 9 | 30 | 10 | 50% |

Table VIII. Computational parameters of energy scheduling models

| Models | SCUC | NNBD-SCUC (0% Sparsity) | SNNBD-SCUC (50% Sparsity) |
|---|---|---|---|
| Objectives | 1 | 1 | 1 |
| # Constraints | 8,086 | 27,646 | 27,646 |
| # Binary Variables | 1,584 | 5,424 | 5,424 |
| # Continuous Variables | 2,401 | 10,441 | 10,441 |
| # Non-zero Coefficients | 16,321 | 84,106 | 68,626 |

As shown in Table VIII, the SNNBD-SCUC model encapsulates the integration of battery degradation within the SCUC optimization framework. In comparison to the conventional SCUC model, notable increases are observed in the numbers of constraints, binary and continuous variables, as well as non-zero coefficients for the 0%-sparsity SNNBD-SCUC model, consequently heightening computational complexity. However, the 50%-sparsity SNNBD-SCUC model exhibits a noteworthy 18.4% reduction in non-zero coefficients, potentially alleviating computational burden and enhancing efficiency.

The outcome for the IEEE 24-bus system with different sparsity levels of the SNNBD model are presented in Table IX. It is vital to recognize that all tabulated results are anchored on a relative MipGap of 0.001, which is a critical gauge of the optimality gap. The table clearly demonstrates that the solving time decreases exponentially as the sparsity level of the SNNBD model increases. The results based on the 60% and 70% sparsity have not been included as the BESS output curve deviates significantly from the solutions based on lower sparsity level models. The 0% and 10% sparsity models results are not listed here since they cannot be solved within the given time frame, whereas the 50% sparsity model requires only 455 seconds for solution. Similarly, for the 20% sparsity model, the day-ahead scheduling problem cannot find the optimal solution within the span of 20 hours, resulting in a reported non-optimal benchmark result.

Table IX IEEE 24 bus day-ahead scheduling results based on different SNNBD models

| Sparsity Percentage | Operation Cost ($) | Degradation Cost ($) | Updated Total ($) | Time (s) |
|---|---|---|---|---|
| 20% | 259,435 | 9,933 | 269,368 | 72,000 |
| 30% | 259,848 | 10,447 | 270,295 | 4,383 |
| 40% | 260,186 | 9,806 | 269,992 | 1,858 |
| 50% | 259,472 | 6,789 | 266,261 | 455 |

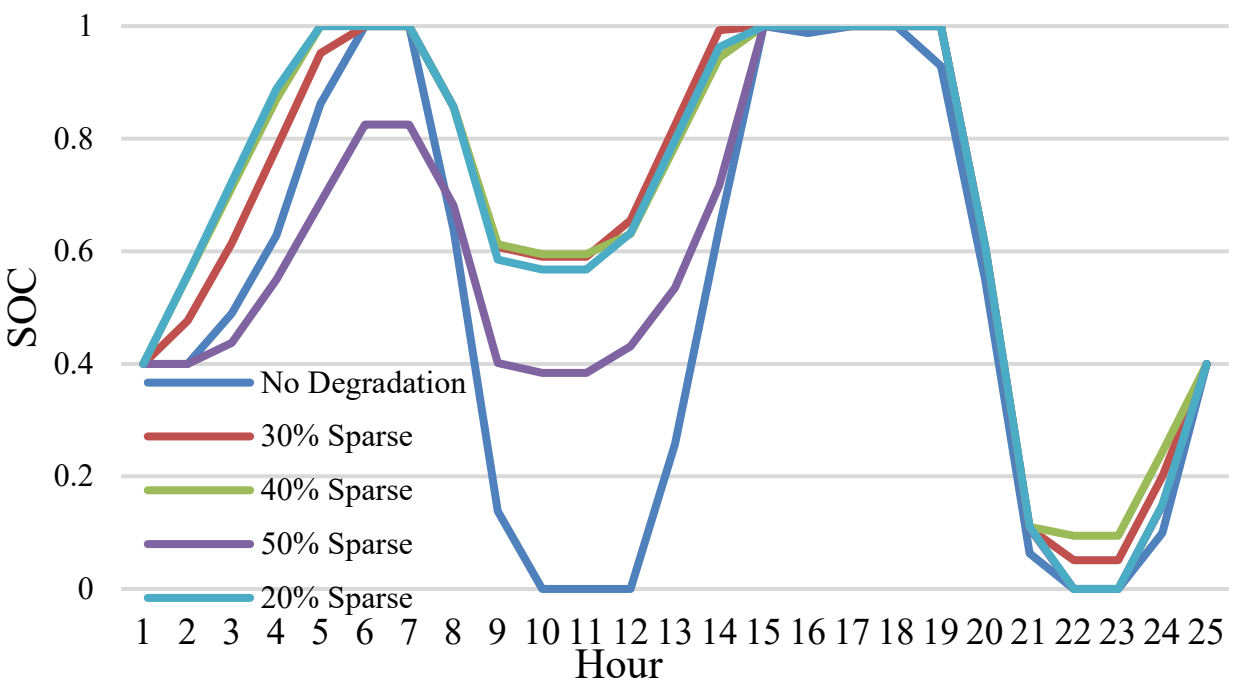

Fig. 6. SOC curves of BESS #4 in the 24-bus bulk power system.

We also found that the 50% sparsity model lead to the minimum total cost. However, the total cost, does not change significantly despite the variation in solving time. This indicates that while the solving time is reduced to an acceptable number with high sparsity SNNBD models, the overall cost remains relatively stable. By analyzing these results, it becomes evident that increasing the sparsity level in the SNNBD model significantly reduces solving time without significantly impacting the overall cost. However, it is crucial to validate the BESS power output pattern when assessing the performance of the SNNBD model. Examining the BESS power output pattern ensures that the model captures the desired behavior and produces outputs consistent with expectations.

Figs. 6 display the SOC curves of BESS #4 under different sparsity levels of the proposed SNNBD model. BESS #4 is particularly active among the five units considered in the testbed. For benchmarking purposes, the SOC curve is also plotted when there is no battery degradation considered in the day-ahead scheduling problem. The SOC curve provides insights into the utilization of the BESS, with more fluctuation indicating more active and flatter curves indicating less active. When degradation is not considered, the BESS units are utilized to their maximum capacity since there is no equivalent degradation cost factored into the optimization problem. We found that both BESS #4 are scheduled to discharge to 0% SOC twice when degradation is not considered. The output curves of BESS #4 with SNNBD models significantly shrink compared to the case where degradation is not considered. However, the output patterns of BESS #4 with different sparsity levels of the SNNBD model exhibit a similar pattern and overlap for most time periods, which demonstrating the effectiveness of the proposed SNNBD model.

Table IX provides insights into the tradeoff between sparsity and accuracy. A higher sparsity level leads to lower accuracy, while a lower sparsity level results in longer solving times for day-ahead scheduling. Thus, a balance must be compromised between sparsity and accuracy. Overall, the 50% sparsity model performs the best since its SOC curve closely resembles those of the 20%, 30%, and 40% sparsity models while having the lowest total cost.

### E. *Market Analysis*

Fig. 7 presents sample results demonstrating the influence of locational marginal price (LMP) when integrating BESSs into the bulk power system. The LMPs are derived from a subsequent Economic Dispatch (ED) run, in which the unit commitment statuses obtained from the SNNBD-SCUC model are fixed. Specifically, the process follows two main steps: (i) the SNNBD-SCUC model first determines the unit commitment and dispatch schedule, and (ii) the fixed-commitment ED model is then solved to obtain the LMPs from the dual variables associated with the nodal power balance constraints. This two-stage procedure ensures that the price calculation remains consistent with standard SCUC practice, and that the embedded neural network and mixed-integer linear programming components within the SNNBD formulation do not directly influence the dual-based price formation. Our exploration encompassed 3 models including “no BESS model”, “BESS considered with degradation”, and “BESS considered without degradation”. A comparison was made with the "no BESS model" to assess the system's ability to reduce line congestion when a BESS is integrated. The LMP results in Fig. 7 specifically focus on bus 14, the location of the largest BESS unit, BESS #4. From the figure, it is evident that during most time periods, such as 1 am to 5 am and 12 pm to 6 pm, the LMP values are consistent across the different cases, indicating there is no line congestion at bus 14 during those periods. However, as the clock strikes 3 pm to 6 pm, a surge in LMP is evident, which suggests that the line is loaded higher than in the previous hours but is not yet congested. During the normal daily peak load periods of 7 am to 9 am and 7 pm to 8 pm, the LMP values differ among the proposed models. In comparison to the "no BESS model," the models with integrated BESS units can significantly reduce the LMP, indicating that the BESS can alleviate line congestion. Note that when battery degradation is not considered, the BESS exhibits a higher capability to mitigate line congestion, leading to the lowest LMP during those congested hours. This analysis of LMP with the integration of a BESS system provides valuable insights for both grid operators and BESS investors, as BESS installations play a crucial role in addressing line congestion within the grid. It is worth to mention that the SNNBD-SCUC does not directly modify the pricing formulation, as LMPs are obtained from a subsequent ED run with fixed unit statuses. Nevertheless, the SNNBD model affects dispatch behavior by capturing nonlinear degradation costs, which influence charging/discharging schedules and alter congestion patterns. This leads to indirect but interpretable changes in LMP values compared with conventional SCUC outcomes.

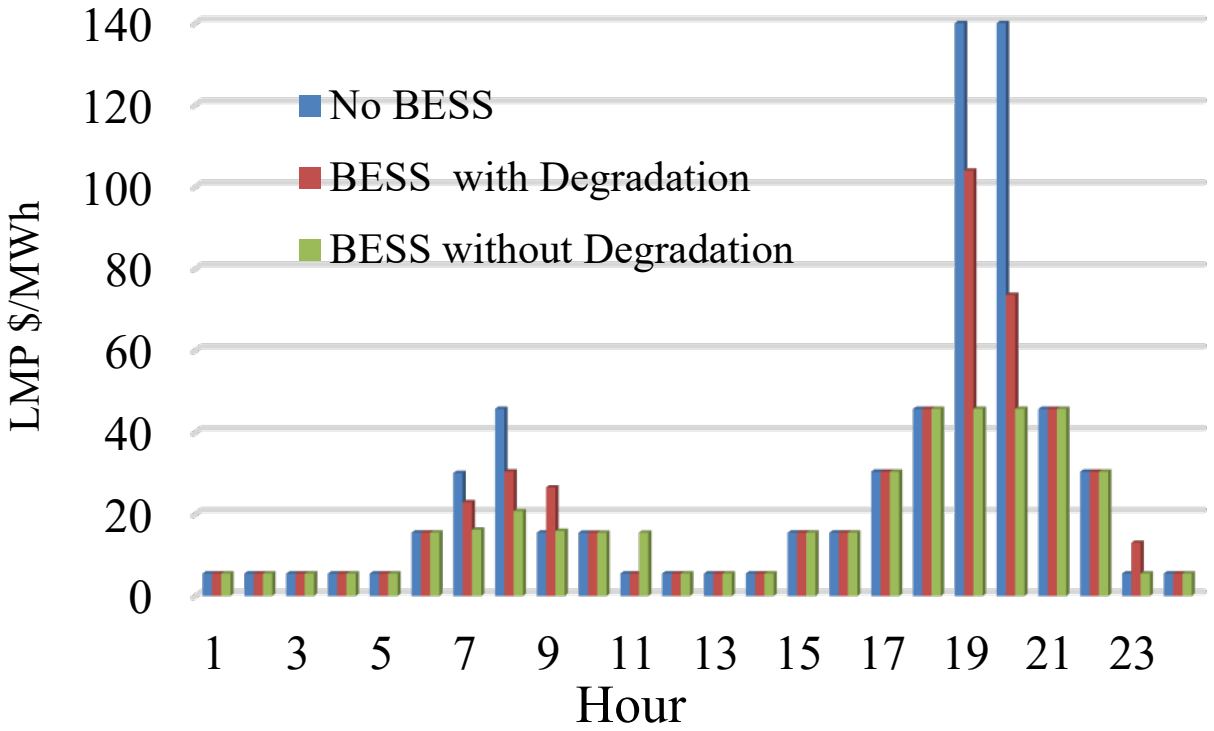

Fig. 7. LMP at bus 14 (BESS #4).

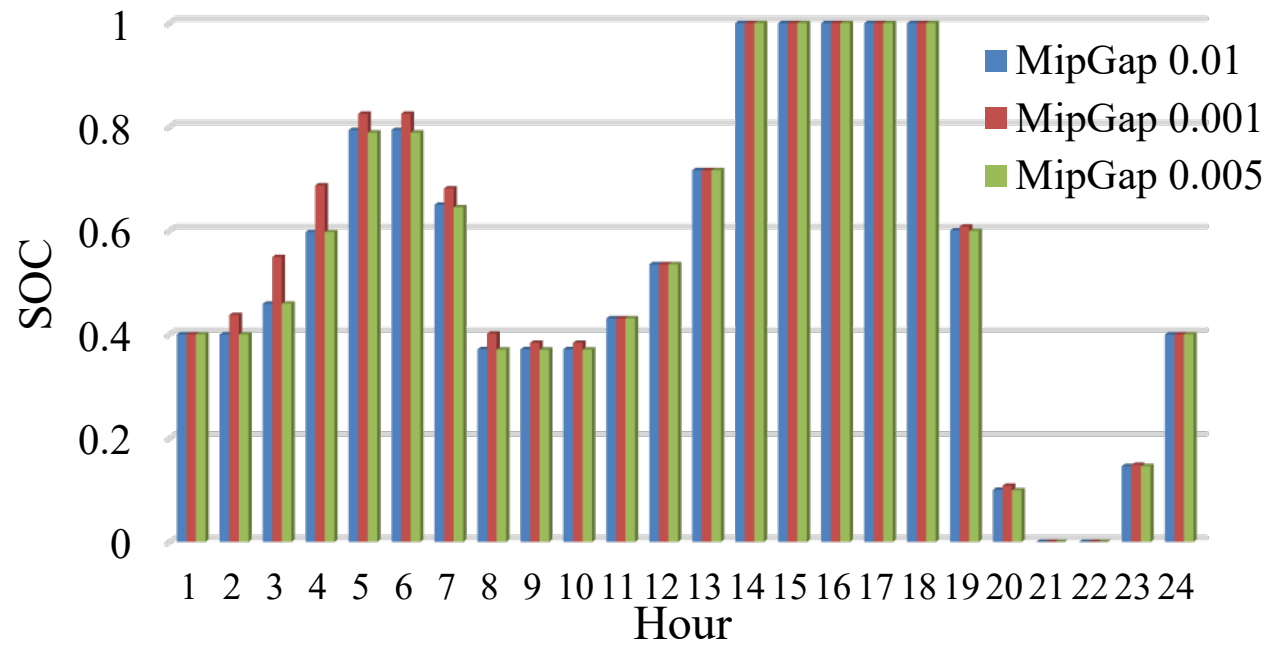

Fig. 8. Scheduled SOC levels of BESS #4 at bus 14.

### F. *Sensitivity Analysis of Relative Optimization Gaps*

A sensitivity test was conducted to examine the impact of different relative gaps. Fig. 8 displays SOC curves of BESS #4 based on the optimal solution obtained using different relative MipGap values. The results presented in Fig. 8 are based on the 50% sparsity SNNBD model. The solving times for MipGap values of 0.01, 0.005, and 0.001 are 339 seconds, 380 seconds, and 450 seconds, respectively. The solving time increases as the MipGap value decreases because a more accurate optimal solution is sought. However, upon analyzing the SOC curve depicted in Fig. 8, it becomes evident that the SOC curves mostly overlap, indicating minimal differences between the solutions obtained under different MipGap values. Consequently, for the 50% sparsity model, a higher MipGap value is preferred as it reduces the computation time while maintaining a comparable solution quality.

## VI. Conclusions

This paper introduces a novel sparse neural network-based battery degradation model that can accurately quantify battery degradation in BESS and largely address the computational challenges associated with traditional dense neural networks when being incorporated into optimization models. By leveraging the sparse technique, the proposed SNNBD achieves accurate degradation prediction while significantly reducing the computational complexity. It has been proven that the accuracy of the SNNBD model does not decrease significantly until the sparsity is increased to 80%. It can obtain 91% accuracy at 60% sparsity, compared to 95% when no sparsity is implemented.

The results also show that our proposed SNNBD model can significantly reduce the computational burden, making neural network-integrated day-ahead energy scheduling directly solvable, even for complicated multi-BESS integrated systems. Furthermore, the results have been proven to be accurate and feasible with a high sparsity SNNBD model in both microgrid and bulk power system. Choosing different sparsity levels for the proposed SNNBD model provides flexibility for the grid operator, as it involves a tradeoff between accuracy and solving time. Overall, the SNNBD model opens up new possibilities for efficiently addressing battery degradation in day-ahead energy scheduling for multi-BESS systems.

## VII. Future Work

A promising direction for future work is to explore the integration of DNN sparsity with other computational efficiency techniques in MILP-based scheduling. For example, sparse neural networks can be combined with decomposition methods such as Benders' decomposition, which partition large-scale problems into smaller, tractable subproblems, or with recent machine learning-based size reduction approaches that predict a subset of decision variables or non-binding constraints with high confidence and remove them from the optimization model. Investigating how these methods can collectively reduce computational complexity while preserving solution quality would provide valuable insights for large-scale energy storage scheduling problems. Such research could enable faster and more scalable integration of detailed battery degradation models into practical day-ahead and real-time power system operations.